\documentclass[5p]{elsarticle}

\usepackage{amsmath}

\usepackage{graphicx}
\usepackage{subfig}
\usepackage{booktabs}

\usepackage{amssymb}
\usepackage{makecell}

\usepackage{lineno,hyperref}
\modulolinenumbers[5]

\journal{Electrochemica Acta}

\DeclareUnicodeCharacter{2061}{}

\biboptions{longnamesfirst, semicolon, sort&compress}

\begin{document}

\begin{frontmatter}

\title{Towards Rigorous Validation of Li-S Battery Models}

\author[mymainaddress]{Cornish,M.\corref{mycorrespondingauthor}}
\cortext[mycorrespondingauthor]{Corresponding author}
\ead{m.cornish14@imperial.ac.uk}

\author[mymainaddress]{Marinescu, M.}

\address[mymainaddress]{Department of Mechanical Engineering, Imperial College London, SW7 2AZ, United Kingdom}

\begin{abstract}
Achieving Li-S batteries' promise of significantly higher gravimetric energy density and lower cost than Li-ion batteries requires researchers to delineate the most important factors affecting the performance of this technology. By encoding this knowledge into a mathematical model, understanding is made precise, quantitative, and predictive. However, the complex and unknown mechanisms of Li-S batteries have multiple proposed models with relatively few informative quantitative comparisons to experimental data. Consequently, a minimum set of testing procedures for model validation is proposed. Moreover, in the absence of an accepted standard model, a zero dimensional model is proposed in this work. The model improves upon several existing models while remaining as simple as possible. The model is quantitatively predictive, as demonstrated by out-of-sample predictions of experimental discharge resistance. Finally, this model and others have been implemented using PyBaMM. Therefore, the open access code allows rapid modifications of this model by all researchers.

\end{abstract}


\end{frontmatter}


\section{Introduction}\label{sec:intro}
Lithium Sulfur (Li-S) battery technology is a promising beyond-Li-ion battery candidate. With a theoretical specific energy of 2567 Wh kg$^{-1}$, Li-S batteries claim an order of magnitude improvement upon Li-ion cells \cite{bruce2012}. Moreover, the materials required for Li-S cells can be less detrimental to the environment \cite{song2013}. Despite being demonstrated in space exploration \cite{EuroSpace} and as a cold temperature energy delivery system \cite{LowTemp}, the development of practical Li-S batteries is still hindered by the technology's complex mechanisms which are not fully understood. Without a sufficient theoretical understanding of these dynamics, the effects of changes to cell design, materials, and operation are difficult to predict, time consuming to discover, and thus costly to optimise. 

Mathematical models can shorten the development cycle of technology through optimisation of design and operation \cite{becker2005}. For example, a model can illustrate the ideal case of a new material design, such as mesoporous cathode particles which trap polysulfides \cite{Danner2015}, and therefore can inform the research community of the maximum benefit associated with such new materials. However, due to the complex nature of Li-S cells, a model which can reliably inform optimisation decisions has not been established. To understand why no model is fully trusted to optimise cell design or operation, an understanding of the success criteria of a model and its use-case is needed.  

The application of Li-S battery models to the optimisation of this battery technology can be categorised as use in a Battery Management System (BMS) and in both cell design and material choice optimisation. Models intended for a BMS are only required to accurately predict observable states, such as voltage and resistance, under new operational loads; otherwise a simple look-up table would suffice. Models intended to inform cell design and material choice must accurately predict cell states for untested cell designs operated under a variety of loads. In this way, the requirements of models in a BMS are a subset of those for cell design. Trust in a model for a specific application requires evidence that the model can perform as intended. However, as is discussed in section \ref{sec:lit_testing}, the models available in the literature have insufficient evidence towards use in BMS and, consequently, design use-cases. 

Establishing trust in a model, or simply model validation, can be established experimentally by simulating the use-case scenario. A BMS model is constructed under a subset of all possible operational loads and required to accurately predict the cell behaviour during unobserved operational loads. Researchers can simulate this use-case scenario by restricting the parameterisation procedure of the model to the experimental data from a subset of operational loads. The data used for fitting is termed \textit{in-sample} and the remaining data is termed \textit{out-of-sample}. Predictive accuracy is tested by comparing the model output against the out-of-sample data. This experimental model validation technique is successfully applied in machine learning and statistics \cite{hastie2009elements}. All battery models, including physics-based models, can benefit from this added rigour in testing to provide evidence that the model is useful in real-world scenarios.  

Unlike strictly data-driven models, the mathematical form of battery models can be physically informed. Such models not only allow a physical interpretation of the mathematical terms involved but also are likely to generalise well from in-sample to out-of-sample data because the terms in the model come from a well established physical principle or relationship. For this reason, we restrict our attention to physics-based models. However, model parameters which are fit to experimental data aren’t guaranteed to be physically meaningful. While evidence for a model's predictive power can be provided by out-of-sample testing, additional testing procedures are required to validate if the model corresponds meaningfully to the physical cell. Such techniques are beyond the scope of this work and we focus our attention to the experimental validation technique discussed above. The purpose of this work is to highlight the problems associated with the model validation techniques currently found in the Li-S modelling literature, to outline a more rigorous framework for testing which will also allow clearer comparisons between models, and to provide a novel zero-dimensional model. The novel model is rigorously validated and provides quantitatively accurate predictions of Ohmic resistance. 

In section \ref{sec:min_tests}, we provide a set of experimental tests based on specific operational loads and observable variables which most proposed Li-S cell-level physics-based models are capable of producing. Thereafter, we consider in section \ref{sec:lit_testing} the experimental validation technique and the consequences of its absence throughout the literature. In particular, we find that Ohmic resistance is largely neglected despite its importance. Next, we propose a new zero-dimensional model capable of predicting accurate out-of-sample resistance data. The model is outlined in section \ref{sec:model_dev} and validated in section \ref{sec:model_val}. By providing both the successes and failures of the model, researchers will have further insight into the necessary modifications of the model along with a clear demonstration of the testing procedure. Closing remarks are given in section \ref{sec:conclusion}

This work is part of a larger effort to increase rigour in the model validation process of Li-S battery models, as outlined in section 14 of \cite{Robinson2021}, as well as a push for greater openness and transparency in the scientific community. As such, the model developed here, along with those found in \cite{Marinescu2016}, \cite{Marinescu2018}, and \cite{Hua2019}, have been implemented in PyBaMM \cite{Sulzer2020} and made publicly available concurrently with this publication \cite{Cornish_LiS-Models_2021}.

\section{Minimum Simulation Test Set}\label{sec:min_tests}

The minimum test set is both a technique to avoid over-fitting as well as a tool of comparison between models. \textit{Over-fitting} occurs when a model has unknown parameters that are fit to data such that the model can adequately reproduce the in-sample data but not the out-of-sample data. Often, reducing model complexity decreases in-sample accuracy but improves out-of-sample predictions. For example, a battery model may accurately fit to experimental voltage data from a constant current discharge and fail to accurately predict the subsequent constant current charge voltage data; a simpler model can achieve both voltage curves, but less accurately overall. A discussion of this exact scenario in the Li-S modelling literature, among other similar cases, is discussed in section \ref{sec:lit_testing}. Comparisons between models is a more involved process.  

Given the difficulty of comparing models which have each been fit to cells of different designs, along with the necessity of avoiding over-fitting, we propose a list of simulation procedures and the expected evolution of macroscopic variables that should be included in model validation. The set of operating conditions were chosen in order to capture common Li-S cell behaviours and to use measurable quantities that most physics-based cell-level Li-S models can produce. Therefore, the majority of cells used for modelling and the models themselves can utilise this test set. In the case that a limited set of experiments for a cell does not consider all operational loads in the test set, the model validation can utilise the expected behaviours given. Indeed, the model validation in section \ref{sec:model_val} does just that. The test set is summarised in Table \ref{tab:tests} along with references which contain experimental work illustrating the cell behaviour. 

\begin{table*}
\begin{center}
\caption{A minimum set of tests for cell-level Li-S battery models along with the expected behaviour of macroscopic variables. This set is derived from available experimental data and will require modification with further experimental findings. Typical Li-S batteries are expected to output these behaviours under the specified operational loads. Simulating these operations out-of-sample should be considered a part of model validation.}
\label{tab:tests}
\begin{tabular}{clc}
\hline
Operational Load & Expected Behaviour(s) & References \\
\hline    
Constant Current Discharge & Two plateau voltage curve, & \cite{Mikhaylik2004, Kumaresan2008, Hofmann2014, Danner2015, Ren2016, Yoo2015, Zhang2015, Zhang2016, Andrei2018, Marinescu2016} \\ 
 & possible voltage dip during plateau transition, \\
 & resistance peak during voltage plateau transition, & \cite{Zhang2015, Hunt2018}\\
 & and minimum resistance near beginning and end of discharge. \\ \\
Constant Current Charge & Two plateau voltage curve, & \cite{Mikhaylik2004, Ren2016, Yoo2015, Marinescu2016, Xiong2019}\\
 & resistance peak during voltage plateau transition, & \cite{Zhang2016}\\
 &  and minimum resistance near beginning and end of charge. & \cite{Hunt2018} \\ \\
 Discharge at higher C-rates & Decrease in low-plateau capacity, & \cite{Mikhaylik2004, Hofmann2014, Ren2016, Zhang2016, Andrei2018}\\
 & possible decrease of high-plateau capacity, \\
 & change in low-plateau capacity greater than upper plateau, \\
 & voltage curve is lower across discharge, \\
 & and initial voltage changes little. \\ \\
 Charge at higher C-rates & Existence of infinite capacity/voltage bifurcation, & \cite{Mikhaylik2004, Xiong2019, Hua2019}\\
 & lower plateau voltage increases, \\
 & and initial voltage changes little. \\ \\
 Constant Current Cycling & Decreases in charge and discharge capacity. & \cite{Marinescu2018}\\ \\
 Increased Depth-of-Discharge & Voltage kink during charge. & \cite{Yoo2015,Zhang2018, Hua2019}\\
 \hline
\end{tabular}
\end{center}
\end{table*} 

 The first proposed operating condition is a constant current discharge. The expected behaviour of such an experiment is that the discharge voltage curve has two distinct plateaus, commonly referred to as the high and low plateaus. Further, it is often observed that a voltage dip occurs during the high-to-low plateau transition. During constant current discharge, Ohmic resistance typically follows an inverted ``V'' shape, where the peak value of resistance is found during the plateau transition. Local maxima of resistance have also been observed at the beginning and end of discharge, leading to a ``sombrero'' shaped resistance curve. 
 
 The second proposed operating condition is a constant current charge. A two-plateau voltage curve is again expected, however it is not always the case that the plateau transition is very distinct. The resistance curves follow the same general pattern as in the constant current discharge. 

Along with the above two operating conditions, we also propose the behaviours observed across operating conditions. A common observation is that, given two cells discharged at two different constant currents, the high-rate cell will have a decrease in overall capacity compared to the low-rate cell. This decrease is primarily caused by a loss in the low plateau rather than the high plateau capacity. Indeed, the high-plateau capacity tends to be relatively stable for moderate changes to discharge rate. Further, the high-rate cell will typically have lower voltage values across the discharge while the value of the discharge voltage at the beginning of discharge is relatively stable.

The next test considers changes in charge behaviour for different values of constant current rates. The low-plateau voltage often increases with increasing current. Moreover, a bifurcation point for the rate exists, where a lower rate can lead to a seemingly infinite-capacity charge at a nearly constant voltage. A larger current leads to an end-of-voltage spike that exceeds operating limits, which we will refer more simply to as an infinite-voltage spike.

The final two tests we consider are the effects of cycling on the voltage curve and the effects of depth-of-discharge (DoD) on the behaviour of voltage at the beginning of the subsequent charge. Cycling procedures generally lead to a decrease in both discharge and charge capacity. An increased DoD tends to produce a voltage kink, where the initial charge voltage spikes above the low plateau value and then drops back down to the low plateau value.

Although the above described test set is common among most cells, newer designs and materials need to be incorporated. There are two cases to consider: either the physical cell used to fit the model is consistent with the test set or atypical behaviours are observed. In either case, it is important to consider how these operational loads affect the cell as this provides insight into the novelty of the new cell. Model validation should then follow the operational loads provided and use the experimental data as the set of behaviours to capture. 

A brief summary of model validation in the literature with regard to the proposed test set it given in Table \ref{tab:lit_testing}. Each macroscopic output for the operational loads discussed above is considered for a given proposed model. A checkmark indicates that the model was published along with a simulation of the model using the operational load and the specific model output was given. The voltage and Ohmic resistance for constant current discharge, charge, cycling is included, as well as the voltage for multiple discharge and charge rates. Moreover, the effect of depth-of-discharge on charge voltage is included. Table \ref{tab:lit_testing} shows that model testing is both sparse and clustered. The sparsity of the tests indicate that many tests are not published. The clustering of the tests indicate that most model outputs chosen for publication are similar. This clustering is not inherently concerning, indeed similar operation loads and model outputs are needed to compare models. However most models claim to accurately predict the output in question. If many models with different mechanisms and physics all claim to adequately model the same phenomena, then further testing is required to determine which model is the best representation of the cell. Therefore, the clustering coupled with the sparsity of tests presents a limited set of evidence towards model selection. 

\begin{table*}[ht]
\begin{center}
\caption{Overview of testing in the Li-S modelling literature with respect to the minimum test set outlined in Table \ref{tab:tests}. A check implies that the model labelled in the row was published with the model output labelled in the column. Most models are not accompanied with many relevant simulations. Moreover, many models present the same tests. No consideration of in-sample and out-of-sample testing is included in the table. The model outputs from left to right are \textit{Discharge Voltage}, \textit{Charge Voltage}, \textit{Cycling Voltage}, \textit{Discharge Ohmic Resistance}, \textit{Charge Ohmic Resistance}, \textit{Discharge Rate-Capacity Relationship}, \textit{Charge Rate Dependent Infinity Voltage/Capacity Effect}, and \textit{Depth-of-Discharge Dependent Initial Charge Voltage Kink}.}
\label{tab:lit_testing}
\begin{tabular}{|l||m{12mm}|m{12mm}|m{12mm}|m{12mm}|m{12mm}|m{12mm}|m{12mm}|m{12mm}|m{12mm}}
\hline 
\enskip & \multicolumn{8}{c|}{Model Output}\\
\hline
Reference & Dis V & Cha V & Cycle V & Dis $\Omega$ & Cha $\Omega$ & Dis Rate-Capacity & Cha Infinite V/Ah & DoD Cha Kink \\
\hline
Mikhaylik \& Akridge \cite{Mikhaylik2004} & \enskip & \checkmark & \enskip & \enskip & \enskip & \enskip & \checkmark & \enskip \\ \hline  
Kumaresan et al. \cite{Kumaresan2008} & \checkmark & \enskip & \enskip & \enskip & \enskip & \enskip & \enskip & \enskip \\ \hline   
Neidhardt et al. \cite{Neidhardt2012} & \checkmark & \enskip & \enskip & \enskip & \enskip & \enskip & \enskip & \enskip \\ \hline
Fronczek et al. \cite{Fronczek2013} & \checkmark & \checkmark & \enskip & \enskip & \enskip & \checkmark & \enskip & \enskip \\ \hline
Hofmann et al. \cite{Hofmann2014} & \checkmark & \checkmark & \checkmark & \enskip & \enskip & \checkmark & \checkmark & \enskip \\ \hline
Zhang et al. \cite{Zhang2015} & \checkmark & \enskip & \enskip & \checkmark & \enskip & \checkmark & \enskip & \enskip \\ \hline
Danner et al. \cite{Danner2015} & \checkmark & \checkmark & \checkmark & \enskip & \enskip & \enskip & \enskip & \enskip \\ \hline
Marinescu et al. \cite{Marinescu2016} & \checkmark & \checkmark & \enskip & \enskip & \enskip & \enskip & \enskip & \enskip \\ \hline
Thangavel et al. \cite{Thangavel2016} & \checkmark & \enskip & \enskip & \enskip & \enskip & \checkmark & \enskip & \enskip \\ \hline
Barai et al. \cite{Barai2016} & \checkmark & \enskip & \enskip & \checkmark & \enskip & \checkmark & \enskip & \enskip \\ \hline
Ren et al. \cite{Ren2016} & \checkmark & \enskip & \enskip & \enskip & \enskip & \checkmark & \enskip & \enskip \\ \hline
Yoo et al. \cite{Yoo2015} & \checkmark & \checkmark & \checkmark & \enskip & \enskip & \enskip & \enskip & \enskip \\ \hline
Zhang et al. \cite{Zhang2016} & \checkmark & \enskip & \enskip & \enskip & \enskip & \checkmark & \enskip & \enskip \\ \hline
Marinescu et al. \cite{Marinescu2018} & \checkmark & \checkmark & \checkmark & \enskip & \enskip & \enskip & \enskip & \enskip \\ \hline
Andrei et al. \cite{Andrei2018} & \checkmark & \enskip & \enskip & \enskip & \enskip & \checkmark & \enskip & \enskip \\ \hline
Hua et al. \cite{Hua2019} & \enskip & \checkmark & \enskip & \enskip & \enskip & \enskip & \checkmark & \enskip \\ \hline
Danner \& Latz \cite{Danner2019} & \checkmark & \checkmark & \checkmark & \enskip & \enskip & \checkmark & \enskip & \checkmark \\ \hline
Xiong et al. \cite{Xiong2019} & \enskip & \checkmark & \enskip & \enskip & \enskip & \enskip & \enskip & \enskip \\ \hline
Erisen et al. \cite{Erisen2018} & \enskip & \enskip & \enskip & \enskip & \enskip & \enskip & \enskip & \enskip \\ \hline
Emerce \& Eroglu \cite{Emerce2019} & \enskip & \enskip & \enskip & \enskip & \enskip & \enskip & \enskip & \enskip \\ \hline
Kamyab et al. \cite{Kamyab2020} & \checkmark & \checkmark & \checkmark & \enskip & \enskip & \enskip & \enskip & \enskip \\ \hline
Parke et al. \cite{Parke2020} & \checkmark & \enskip & \enskip & \enskip & \enskip & \checkmark & \enskip & \enskip \\ \hline
Erisen \& Eroglu \cite{Erisen2020} & \checkmark & \enskip & \enskip & \enskip & \enskip & \enskip & \enskip & \enskip \\ \hline
Parke et al. \cite{Parke2021} & \checkmark & \enskip & \enskip & \enskip & \enskip & \checkmark & \enskip & \enskip \\ \hline

\end{tabular}
\end{center}
\end{table*}  

By following the test set, model validation and selection becomes more rigorous and transparent than it is currently. Within model validation, models which perform poorly under particular operational loads can be easily detected and the necessary modifications can be quickly proposed. Within model selection, the comparison of two models is difficult when one model is published with discharge simulations only and the other with charge simulations only. As discussed in section \ref{sec:lit_testing} and illustrated in Table \ref{tab:lit_testing}, this is not unusual. The limited set of tests provided in the literature provides little information regarding performance when comparing multiple models. \

\section{Testing in the Literature}\label{sec:lit_testing}

Experimental validation, as discussed in sections \ref{sec:intro} and \ref{sec:min_tests}, is an essential step in model development. The absence of this testing procedure can mislead researchers as to the quality of a model. Without open-source simulation code available, the absence of this validation procedure requires researchers to re-implement the model in order to complete the tests. This necessarily reduces time for original research. Open source code is not frequently published and so researchers must rely on the published testing. Throughout the literature it is common to validate the model by only considering in-sample simulations, either with or without experiment data to compare against. Out-of-sample simulations, if they exist, are rarely compared against experimental data. Moreover, out-of-sample predictions are often in the form of variables internal to the cell for which the relevant experimental data is neither available nor incorporated. A clear example of the problems associated with limited model testing can be seen through the highly influential model from Kumaresan et al. \cite{Kumaresan2008}.

The Kumaresan model is the first spatially resolved cell-level model for Li-S batteries. This model presents a highly detailed system, requiring many parameters to fit. The parameter set was fit to a single C/50 discharge voltage curve. Therefore, any other operational load can be considered out-of-sample. However, without the original data set, or a very similar cell to test, it remains difficult to compare out-of-sample predictions directly to experimental values. Nonetheless, we can observe how the model performs compared to the typical behaviours from the standard test set detailed in section \ref{sec:min_tests}.

Ghaznavi \& Chen \cite{Ghaznavi2014a} performed a sensitivity analysis of the Kumaresan et al. \cite{Kumaresan2008} model. Along with analysis of the  parameters, Ghaznavi \& Chen simulated discharge voltage curves for a range of currents from 0.2C to 7C. The typical low-plateau capacity loss associated with high currents is not observed until, approximately, the 7C discharge. The capacity does decrease with increasing currents; however, this is the result of the high-plateau capacity loss. Moreover, there is a pronounced downward shift across the entire voltage curve, which is not typical of Li-S cells with higher C-rates. Further analysis from Ghaznavi \& Chen \cite{Ghaznavi2014b} revealed that the model cannot charge without significant change to the value of the saturation concentration of $\text{Li}_2\text{S}_{(s)}$. Although the model is physically reasonable, we observe less reasonable behaviour from these out-of-sample simulations. 

Zhang et al. \cite{Zhang2015} considered the electrolyte resistance simulation from a simplified zero-dimensional model and that from the Kumaresan et al. model. At the low discharge rate of C/50, the IR drop is not expected to have a noticeable effect on the discharge voltage curve. However, the value of the electrolyte resistance can still be calculated from the model and is expected to follow, roughly, an inverted ``V'' shape, with the peak resistance occurring during the voltage plateau transition. The simplified model from Zhang et al. accurately captures the electrolyte resistance curve, but the validation of their model is only in-sample. The Kumaresan et al. model, in contrast, provides a local maximum in resistance during the voltage plateau transition, but the initial and final stages of discharge yield much higher values. As opposed to an inverted ``V'' shape, the model gives a ``W'' shape. This electrolyte resistance curve is not categorically different from other resistance curves. For example, at lower or higher temperatures Hunt et al. \cite{Hunt2018} find smaller local maxima at the start and end of discharge. However, the global maximum of electrolyte resistance consistently appears during the plateau transition, contradicting the Kumaresan et al. model. Finally, the quantitative value of the simulated electrolyte resistance was several orders of magnitude less than expected. Therefore, the effect of the IR drop is unlikely to be captured. 

The Kumaresan et al. \cite{Kumaresan2008} model is illustrative of the difficulties encountered without rigorous testing procedures, but it is not a unique case of minimal testing in the literature. Prior to this one-dimensional model, Mikhaylik \& Akridge \cite{Mikhaylik2004} developed the first cell-level physics-based model for Li-S batteries. The model illustrated a possible, and now widely accepted, causal connection between the polysulfide shuttle phenomena, self-heating, and the end-of-charge infinite-capacity/infinite-voltage bifurcation. The model and subsequent analysis matched the data presented, but it appears the comparison was performed entirely in-sample. As illustrated with the Kumaresan et al. model, the absence of out-of-sample testing can mask large problems which can provide clear evidence against the model's validity. 

Soon after Kumaresan et al. \cite{Kumaresan2008} introduced the first one-dimensional model, other formulations began to emerge. Fronczek \& Bessler \cite{Fronczek2013} simplified the Kumaresan et al. model considerably, but produced atypical discharge curves for different constant current C-rates. It is unknown if the experimental cell used to fit the model also had this behaviour. Without data for comparison, it is difficult to assess the model accuracy. However, this work is unique in the physics-based cell-level modelling literature by producing Electrochemical Impedance Spectroscopy (EIS) predictions. Such EIS predictions could be utilised as an additional testing procedure, because most cell-level physics-based models should be capable of providing such predictions. Zhang et al. \cite{Zhang2015} later note that Fronczek \& Bessler's predictions do not match experimental data. Hofmann et al. \cite{Hofmann2014} greatly advanced the work of Fronczek et al. \cite{Fronczek2013} in a number of ways, including adding electrochemical reactions of polysulfides on the anode. These additional reactions provide an explanation of cycle-life degradation due to passivation of the anode. Moreover, Hofmann et al. provide out-of-sample predictions against experimental data for discharge curves in the range of 0.24C and 1.59C. However, like the Kumaresan et al. model, the high plateau capacities decrease significantly more than the low plateau capacities, unlike the expected behaviour in Table \ref{tab:tests}. The out-of-sample simulations and data presented for this model provide a far clearer picture of model performance than do the published tests of other models. 

Although Hofmann et al. \cite{Hofmann2014} did improve upon the accuracy of discharge voltage curves for different C-rates, as compared to Fronczek \& Bessler \cite{Fronczek2013}, some inaccuracies still remained. Specifically, Fronczek \& Bessler fail to capture the rate-dependent discharge capacity fade, where the low-plateau capacity decreases significantly more than the high-plateau capacity during higher C-rate discharges. Several papers have aimed to explain this capacity fade. Zhang et al. \cite{Zhang2016} argued that transport limitations account for the decreased lower plateau capacity during discharge. Alternatively, Ren et al. \cite{Ren2016} argued the drop in capacity is due to precipitation effects. Andrei et al. \cite{Andrei2018} approached the precipitation model with phenomenological formulations and make similar conclusions to Ren et al. \cite{Ren2016}. Xiong et al. \cite{Xiong2019} enhanced the precipitation hypothesis by incorporating a binary distribution for precipitated particle sizes. None of the models above have been validated with a clear distinction between in-sample and out-of-sample predictions. To distinguish between the multiple hypotheses for the rate-dependent capacity fade, out-of-sample predictive power of a model would present good evidence towards the corresponding hypothesis. Without out-of-sample testing, the predictive power is undeterminable. 

Danner \& Latz \cite{Danner2019} also argued in favour of the precipitation based hypothesis and incorporate out-of-sample predictions of constant current charge and discharge, but for operational loads outside of their data set. Certainly, a lack of data should not halt out-of-sample predictions. Indeed, such predictions are recommended as part of the validation procedure discussed in section \ref{sec:min_tests}. However, this out-of-sample model validation is insufficient given the specific nature of the hypothesis as well as the discrepancy between what data is provided and the model predictions. 

The precipitation-hypothesis papers discussed above also give simulations and data regarding precipitate morphology, possibly leading to an additional testing procedure. However, not all models are capable of producing this data and, as pointed out by Danner \& Latz \cite{Danner2019}, the data is heavily dependent upon cathode design and thus not widely applicable. Given the increasing amount of experimental data now available, many parameters used in these models are experimentally obtained but the large number of remaining fitted parameters associated with each model allows possible over-fitting; a problem which can only be alleviated through testing against data. 

More recently, Kamyab et al. \cite{Kamyab2020} modified the Kumaresan et al. \cite{Kumaresan2008} model to include an active material loss term to account for degradation due to shuttle. A single experimental charge and discharge voltage curve from Yoo et al. \cite{Yoo2015} was directly compared to the simulations. Because the Kamyab et al. model discharge parameters come from Kumaresan et al., which were fit to a different cell at a different C-rate, the Kamyab et al. model discharge voltage curve can be considered out-of-sample. However, the charge parameters were fit to the Yoo et al. experiment and so must be considered in-sample. Moreover, the charge rate-dependent infinite voltage/capacity phenomena associated with polysulfide shuttle was not tested despite the importance of shuttle in the model. Parke et al. \cite{Parke2020} simplified the 1D Kumaresan et al. model via a Tank-in-Series methodology. Although the Tank model matched closely with the 1D model, the discharge rate-capacity effect is not captured. Indeed, the model contradicts typical behaviours by predicting that the upper-plateau capacity decreases for larger currents, while the lower-plateau capacity increases. The net effect is no capacity loss. These rate-dependent predictions are out-of-sample but no experimental data is provided to suggest these results align with a known Li-S cell. Park et al. \cite{Parke2021} improved upon their previous model by incorporating new parameterisations and electrochemical pathways but the distinction between in-sample and out-of-sample was not present. Further, only discharge was considered. Erisen \& Eroglu \cite{Erisen2020} also simplified the Kumaresan et al. model. However, their model produced abnormal discharge voltage curves with a narrow range of discharge voltages and a relatively small high plateau capacity. 

One-dimensional models that have many unknown parameters lead to several concerns, as highlighted above. Such over-fitted models can fail to a capture out-of-sample predictions for charge and discharge. Worse yet, models can fail to charge at all. Several zero-dimensional models have been developed to reduce such overfitting. Marinescu et al. \cite{Marinescu2016} retained the two-stage electrochemistry of Mikhaylik \& Akridge \cite{Mikhaylik2004}, while providing a zero-dimensional dynamical model capable of both discharge and charge. The simulated discharge voltage ranges appear narrower than is typically seen, as can be observed from their in-sample data. In section \ref{sec:model_dev}, we discuss why this model is incapable of retrieving accurate resistance simulations. Indeed, Erisen et al. \cite{Erisen2018} and Emerce \& Eroglu \cite{Emerce2019} use this same electrochemical pathway and fit their model parameters to EIS measurements for 60\% depth-of-discharge. Therefore, the reliability of the parameter fit requires more detailed information than was provided. Moreover, the distinction between in-sample and out-of-sample data is unclear. Marinescu et al. \cite{Marinescu2018} modified their previous zero-dimensional work to include a degradation mechanism due to the shuttle mechanism and anodic passivation. However, the experimental data and simulation comparisons appear to be entirely in-sample. Hua et al. \cite{Hua2019} included thermal effects into the Marinescu et al. \cite{Marinescu2016} model by utilising the same temperature relations from Mikhaylik \& Akridge \cite{Mikhaylik2004}. However, their model validation also appears to be in-sample. As mentioned previously, Zhang et al. \cite{Zhang2015} provide a zero-dimensional model which produces qualitatively correct discharge electrolyte resistance. Barai et al. \cite{Barai2016} combine the zero-dimensional model of Zhang et al., the Butler-Volmer formulation found in Kumaresan et al. \cite{Kumaresan2008}, and a poro-mechanical model for cathode volume changes. The Barai et al. model also accurately produces the discharge electrolyte resistance curve. Both Zhang et al. \cite{Zhang2015} and Barai et al. \cite{Barai2016} do not provide an in-sample/out-of-sample distinction. In section \ref{sec:model_dev}, we develop a zero-dimensional model intended to simplify the Zhang et al. \cite{Zhang2015} and Barai et al. \cite{Barai2016} models, alleviate some of the above discussed issues, and also provide experimental model validation in section \ref{sec:model_val}.

Although typical Li-S cell behaviour is the focus of this work, the recent developments in cathode design and the subsequent modelling efforts need to be considered. Several researchers have modelled cells with cathodes containing meso-porous particles which trap polysulfides in order to limit transport. Such novel cell designs inevitably lead to unexpected behaviours for the macroscopic variables. Therefore, it is essential that out-of-sample predictions are compared to experimental data. Danner et al. \cite{Danner2015} developed a spatially resolved model for a trapped particle cathode in which strong effort is made to utilise experimentally derived parameter values. They provided an in-sample comparison for a single constant current charge and discharge for several cells of different cathode thicknesses. Achieving trust in a model which can predict performance of cells across several design modifications still requires out-of-sample simulations compared against experimental data. However, further considerations are needed for such use-cases and are out-of-scope of this work. Thangavel et al. \cite{Thangavel2016} provided an alternative trapped particle model, with many parameters taken from Kumaresan et al. \cite{Kumaresan2008} and without direct comparisons to data. Yoo et al. \cite{Yoo2015} modified the Kumaresan et al. \cite{Kumaresan2008} model to allow anodic reactions, but only include in-sample validation.

Due to the idiosyncrasies of each Li-S cell design, special care must be made when considering the out-of-sample data used for model validation and comparison of hypotheses. For example, Zhang et al. \cite{Zhang2018} performed an experiment to illustrate their hypothesis that transport mechanisms drive the rate-dependent discharge capacity fade. Zhang et al. \cite{Zhang2016} formulated this transport hypothesis into a model. However, Andrei et al. \cite{Andrei2018} performed the same experiment, found different experimental results, and alternatively hypothesised that precipitation effects are the main mechanism driving the capacity fade. Even if the models were validated against out-of-sample data, these two competing hypotheses cannot be adequately compared across cells which produce different experimental data under the same operational loads. The experimental validation of a model using the procedure outlined in section \ref{sec:min_tests} is limited to cells which produce similar data.  

\begin{figure}[h]
\centering
\subfloat[In-Sample vs Out-of-Sample]{\includegraphics[width=0.25\textwidth]{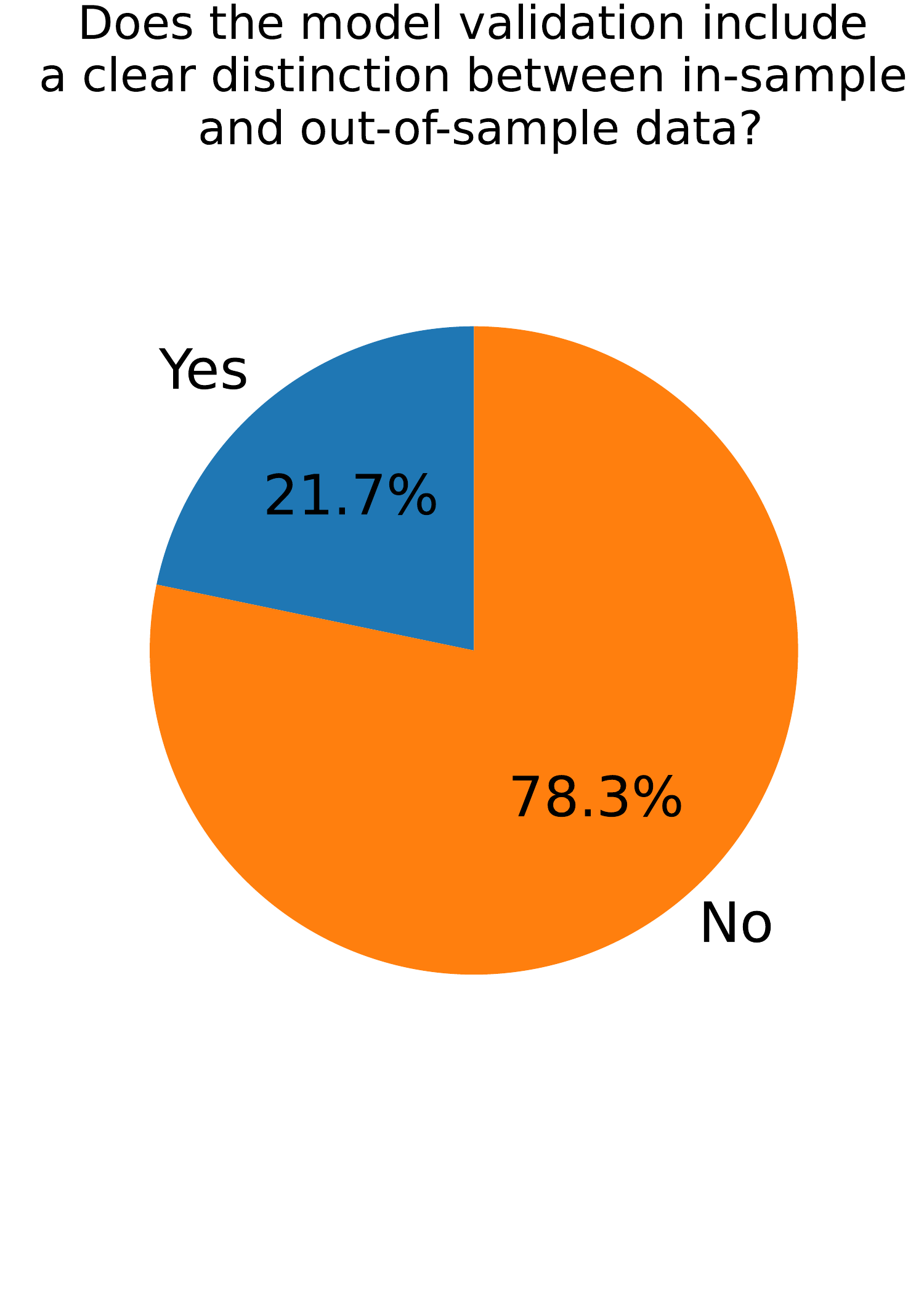}
}
\subfloat[Charge and Discharge]{\includegraphics[width=0.25\textwidth]{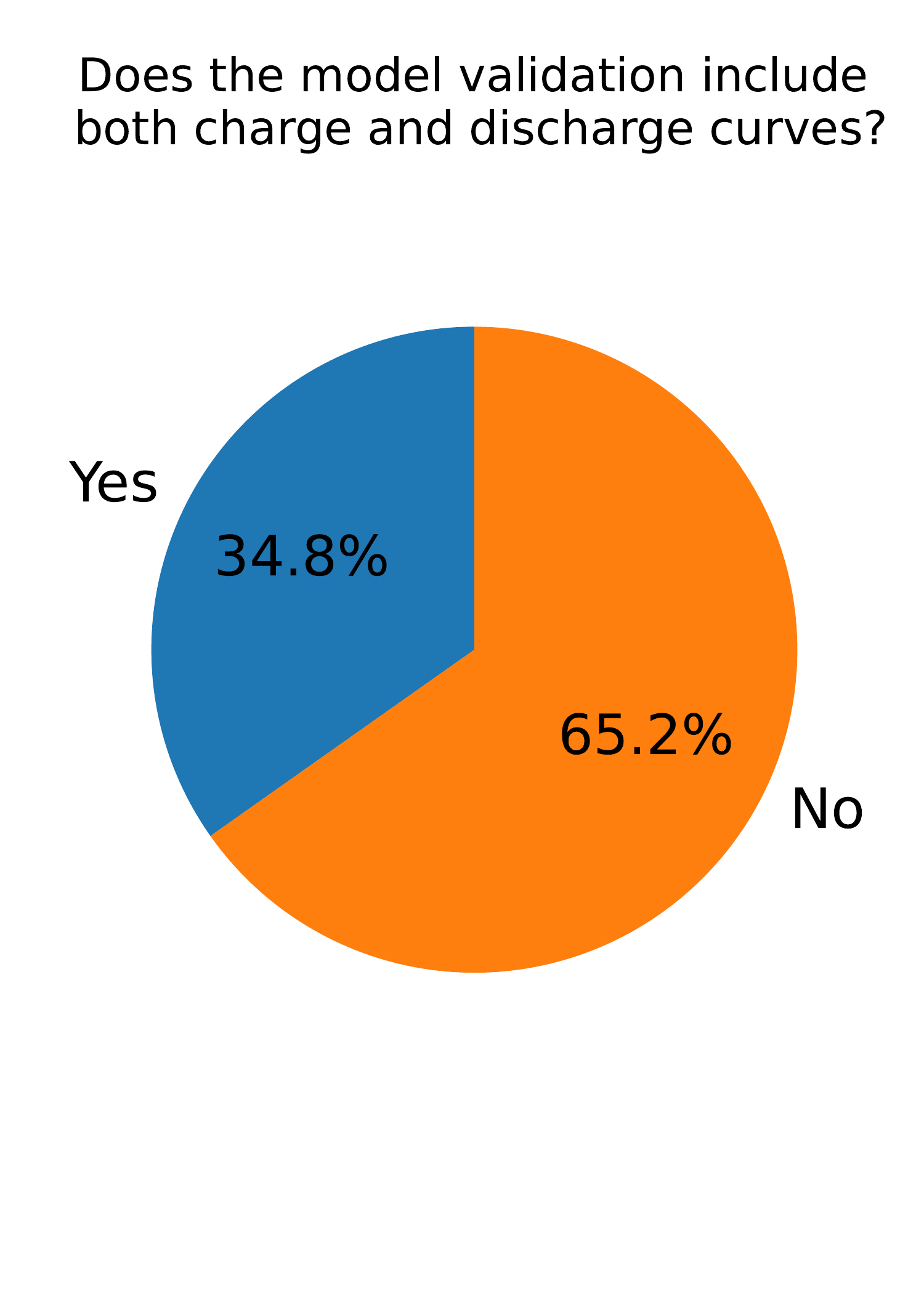}
}    
\caption{(a) Li-S models are not commonly validated with a clear distinction between in-sample and out-of-sample data. The majority of models therefore are not published without clear evidence indicated predictive power. (b) Li-S model simulations commonly only include either discharge or charge simulations. The majority of models therefore are not published with clear evidence indicating the ability to perform both charge and discharge.}\label{fig:lit_graphical_summary}   
\end{figure}

Illustrated in Figure \ref{fig:lit_graphical_summary}, models are rarely published with either out-of-sample validation or both charge and discharge curves. Figure \ref{fig:lit_graphical_summary}(a) presents the portion of models which were published along with a clear distinction between in-sample and out-of-sample model validation. Approximately a fifth of the models are shown to have any out-of-sample validation. Without such tests, the predictive power of the model is unknown. Figure \ref{fig:lit_graphical_summary}(b) presents the portion of models which are demonstrated to be capable of charge and discharge. A little over a third of models demonstrate this ability. Given that the analysis of Ghaznavi \& Chen \cite{Ghaznavi2014b} showed the Kumaresan et al. \cite{Kumaresan2008} model could not charge, it should have become standard practice for new models to consider this problem and provide evidence that such functionality was possible. 

The above discussion reiterates the need for a more standardised and rigorous model validation procedure. Due to the inherent uncertainty in analysing model quality, made especially acute by limited testing, we propose a new minimum model for Li-S cells in section \ref{sec:model_dev} which captures charge and discharge voltage curves as well as accurately predicts out-of-sample Ohmic resistance data. This model also captures variations in these quantities for cells held at different temperatures. Validation of the model is given in section \ref{sec:model_val}.

\section{Model Development}\label{sec:model_dev}

\begin{figure}[h]
\includegraphics[width=0.5\textwidth]{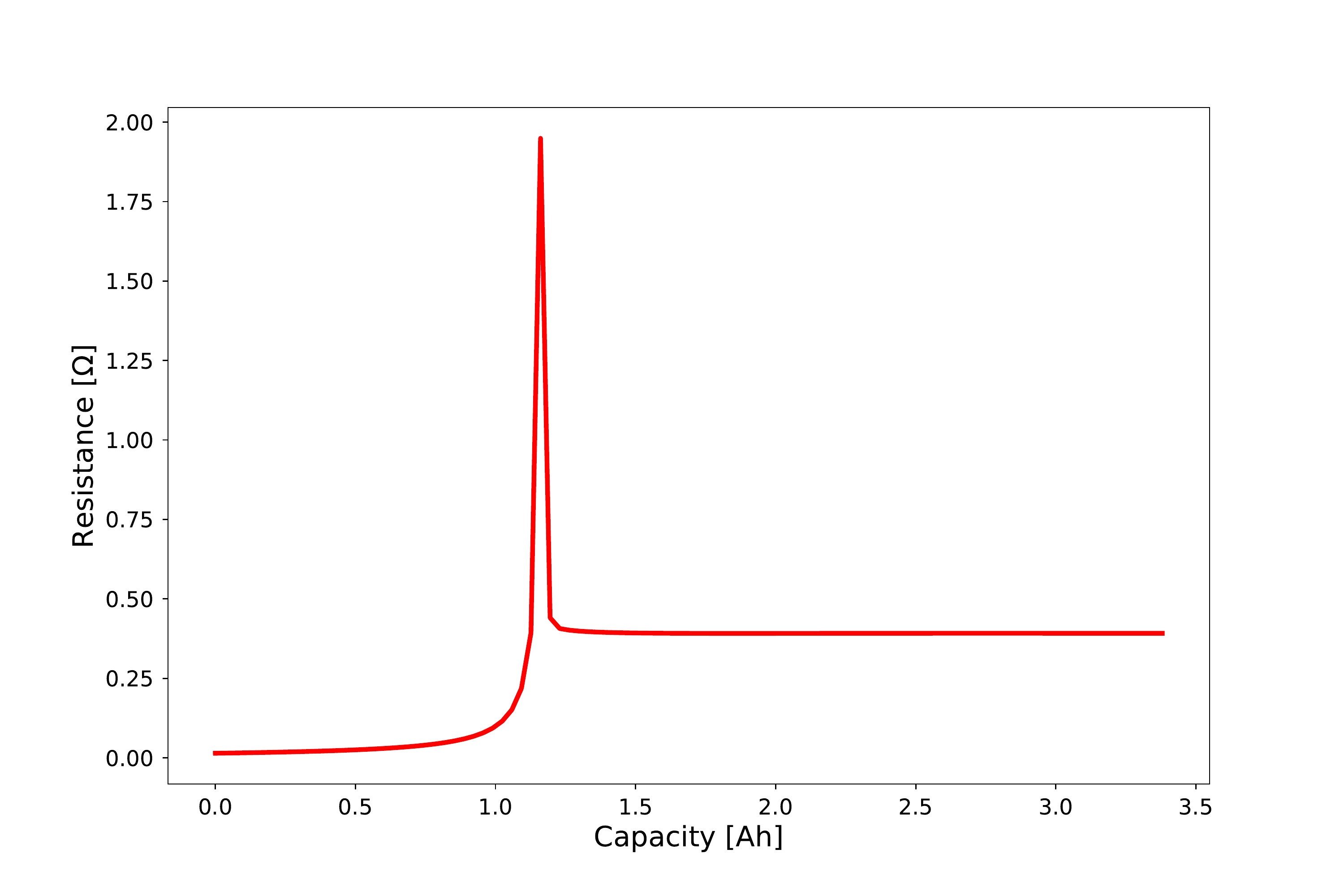}
\caption{Discharge resistance curve found with the Marinescu et al. \cite{Marinescu2016} model for the cell and the resistance model from Zhang et al. \cite{Zhang2015}. The data from Hunt et al. \cite{Hunt2018} shows that the experimentally found resistance curve does not match that of the model.}
\label{fig:marinescu_resistance}
\centering
\end{figure}

The electrochemical pathway given by Marinescu et al. \cite{Marinescu2016} hindered the model from correctly predicting electrolyte resistance. The discharge resistance curve for Marinescu et al. is given in Figure \ref{fig:marinescu_resistance} along with the experimental curve found by Hunt et al. \cite{Hunt2018}. Zhang et al. \cite{Zhang2015} developed the resistance model used. It is assumed that the relation between cell resistance and electrolyte concentration is given through Li cation concentration. The electrolyte resistance is proportional to the reciprocal of the electrolyte conductivity, which in turn is linearly related to the Li cation concentration. Due to charge neutrality, the Li cation concentration can be written in terms of polysulfide anion concentration. During discharge, the two-stage reaction causes higher order polysulfides to reduce and accumulate as both $S_p$, the precipitated sulfur, and $S^{2-}$, the lowest order sulfur anion. The lithium cations increase in concentration as further sulfur anions are produced, but the lithium cation concentration cannot fully decrease after the voltage plateau transition because the accumulated $S^{2-}$ remains in the system to the end of discharge. This electrochemistry is also found in the zero-dimensional models of Marinescu et al. \cite{Marinescu2018} and Hua et al. \cite{Hua2019}.

An improved model is proposed in this section which allows polysulfides to end in precipitated form, and thus the resistance to decrease towards the end of discharge. Further, this model simplifies the zero-dimensional model proposed by Zhang et al. \cite{Zhang2015}, a model which also yielded correct electrolyte resistance behaviour. Thus, the system requires fewer parameters and is less likely overfitted. As a zero-dimensional model, the proposed system simplifies all one-dimensional models. Moreover, the computational cost is low enough for practical application in a BMS.

\paragraph{Mathematical Formulation}

The electrochemical reaction during discharge is modelled as the three-stage reversible reaction,

\begin{subequations}
\begin{align}
S_8^0+4e^- & \rightarrow 2 S_4^{2-}, \\
S_4^{2-}+2e^- & \rightarrow  2 S_2^{2-},\\
S_2^{2-}+2 e^- & \rightarrow  2S^{2-}\downarrow.	
\end{align}
\end{subequations}

Each reaction above, respectively, is referred to as $\text{\it high}$, $\text{\it medium}$, and $\text{\it low}$ and the subsequent mathematical formulations are labelled with the subscripts $H$, $M$, and $L$. These reactions are reversed during charge. 

The reaction rates are given by Butler-Volmer dynamics, 

\begin{subequations}
\begin{align}
i_H &=-2i_H^0 a_r  \sinh⁡\left(\frac{n_H F}{2RT} \eta_H \right), \\
i_M &= -2i_M^0 a_r   \sinh⁡\left(\frac{n_M F}{2RT} \eta_M \right), \\
i_L&= -2i_L^0 a_r   \sinh⁡\left(\frac{n_L F}{2RT} \eta_L \right),
\end{align}
\end{subequations}
where $i_H^0$,  $i_M^0$, and $i_L^0$ are the exchange current densities, $a_r$ is the active surface area of the cathode, $n_H$,  $n_M$, and  $n_L$ are the number of electrons transferred for each reaction, $F$ is Faraday's constant, $R$ is the gas constant, $T$ is the reference temperature of the reactions, and $\eta_H$, $\eta_M$, and $\eta_L$ are the reaction over-potentials on the cathode. The transfer coefficients in the Butler-Volmer equation are assumed to both be equal to 0.5. The active surface area, $a_r$, is assumed constant.

The over-potentials are given as the difference between the electrode voltage, V, and the Nernst potentials $E_H$,  $E_M$, and $E_L$. Therefore, 

\begin{subequations}
\begin{align}
\eta_H &=V-E_H, \\
\eta_M &=V-E_M,\\
\eta_L &=V-E_L.
\end{align}
\end{subequations}

Within the Nernst potentials, Dilute Solution Theory is used to define the activity functions. Therefore, 

\begin{subequations}
\begin{align}
E_H &=E_H^0+\frac{RT}{n_H F}  \ln⁡\left(f_H  \frac{S_8^0}{(S_4^{2-} )^2} \right), \\
E_M &=E_M^0+\frac{RT}{n_M F}  \ln\left( f_M  \frac{S_4^{2-}}{(S_2^{2-})^2} \right), \\
E_L &=E_L^0+\frac{RT}{n_L F}  \ln⁡\left(f_L  \frac{S_2^{2-}}{(S^{2-} )^2} \right),
\end{align}
\end{subequations}
for the standard potentials $E_H^0$, $E_M^0$, and $E_L^0$. The dimensionality constants,

\begin{subequations}
\begin{align}
f_H &= \frac{(n_{S_4^{2-} } )^2 M_S \nu}{n_{S_8^0}} ,\\
f_M &= \frac{(n_{S_2^{2-} } )^2 M_S \nu}{n_{S_4^{2-}}} ,\\
f_L &= \frac{(n_{S^{2-} } )^2 M_S \nu}{n_{S_2^{2-} }} ,
\end{align}
\end{subequations}
are defined for the molar mass of sulfur, $M_S$, the electrolyte volume, $\nu$ and the number of sulfur atoms per molecule $n_{S_8^0 }$, $n_{S_4^{2-} }$, $n_{S_2^{2-} }$, and $n_{S^{2-}}$. 

Along with the electrochemical reaction kinetics at the cathode, the polysulfide induced self-discharge electrochemical reaction on the anode is modelled as an exchange of $S_8^0$ for $S_4^{2-}$ at a rate of $k_s^{(T)} S_8^0$ grams per second, where $k_s^{(T)}$ is referred to as the shuttle constant with the superscript indicating the temperature, $T$, at which the constant is defined. However, the anodic overpotential is expected to be negligibly small and so is not included. Finally, the rate of the precipitation/dissolution reaction is modelled as proportional to the amount of precipitate, $S_p$, multiplied by the difference between $S^{2-}$ and the temperature dependent saturation mass, $S_*^{(T) }$. The precipitation/dissolution rate constant is denoted by $k_{p/d}^{(T) }$. 

The dynamics are governed by the differential algebraic system,

\begin{subequations}
\begin{align}
\frac{dS_8^0}{dt} &= -\frac{n_{S_8^0} M_S}{n_H F} i_H - k_s^{(T) } S_8^0, \\
\frac{dS_4^{2-}}{dt} &= \frac{n_{S_8^0} M_S}{n_H F} i_H + k_s^{(T) } S_8^0-\frac{n_{S_4^{2-} } M_S}{n_M F} i_M ,\\
\frac{dS_2^{2-}}{dt} &=\frac{n_{S_4^{2-} } M_S}{n_M F} i_M -\frac{n_{S_2^{2-}} M_S}{n_L F} i_L, \\
\frac{dS^{2-}}{dt} &= \frac{n_{S_2^{2-}} M_S}{n_L F} i_L - \frac{k_{p/d}^{(T) }}{\nu \rho_S } S_p \left(S^{2-}-S_*^{(T) } \right),\\
\frac{dS_p}{dt} &= \frac{k_{p/d}^{(T) }}{\nu \rho_S } S_p \left(S^{2-}-S_*^{(T) } \right), \\
I &= i_H+i_M+i_L.
\end{align}
\end{subequations}

Finally, resistance is modelled as in Zhang et al. \cite{Zhang2015},

\begin{equation}
	R = \frac{l}{A\left( \sigma_0 - b\sum_iC_i\right)},
\end{equation}
where the summation is taken over the concentration, $C_i$, of anionic species $i\in \{ S_4^{2-}, S_2^{2-}, S^{2-}\}$, $\sigma_0$ sets the maximum resistance when ionic concentration is lowest, and $b$ is a fitted parameter. The data used for model validation is from the experimental work of Hunt et al. \cite{Hunt2018}. 

\paragraph{Fitting Procedure}

As detailed in Table 1 of the supplementary material, several parameters have been assumed. Where suitable, parameters have been fit to the Hunt et al. \cite{Hunt2018} experimental data. This data is derived from a single layer cell held at a constant temperature by a Peltier element and cycled at 0.1C charge and 0.2C discharge. Three constant-temperature cycle experiments were preformed at 20$^{\circ}$C, 30$^{\circ}$C, and 40$^{\circ}$C cell surface temperature. 

Two parameters are assumed to have a binary dependence on current. The precipitation/dissolution rate, $k^{(T)}_{s/p}$ and shuttle constant, $k_s^{(T)}$, are assumed to be different during charge and discharge. All current-independent parameters and charge-specific parameters from the 30$^{\circ}$C charge data have been fit by inspection. The initial condition was found by first assuming the two upper-plateau reaction currents are initially zero, indicating that the initial stages of charge are dominated with dissolution of precipitate sulfur and the low plateau electrochemical reaction. Mathematically, 

\begin{equation}
	i_H=i_M=0.
\end{equation}

Due to the large rate of change of voltage at the start of charge, the initial voltage was chosen to be $V(t=0)=2.0 V$ for numerical stability. This voltage value is near the low plateau voltage value and therefore the rate of change in voltage is far less than would otherwise be at earlier times. Given this voltage value, the equation $i_H=0$ may be rearranged for the initial value of $S_8^0$ in terms of $S_4^{2-}$. Likewise, equation $i_M=0$ may be rearranged for the initial value of $S_4^{2-}$ in terms of $S_2^{2-}$. In particular, 

\begin{subequations}
\begin{align}
S_8^0 &= \frac{\left(S_4^{2-}\right)^2}{f_H^4} \exp\left( \frac{4F}{RT} \left(V-E_H^0 \right)\right),\\
S_4^{2-} &= \frac{\left(S_2^{2-}\right)^2}{f_M^2} \exp\left(\frac{2F}{RT} (V-E_M^0 )\right),
\end{align}
\end{subequations}
where all variables are evaluated at $t=0$.

The initial value $S_2^{2-} (t=0)$ may be found by numerically solving the root finding problem $I=i_L$. In order to compute the root, the value $S^{2-} (t=0)$ is required. This initial value, along with $S_p (t=0)$ were fit to the 30$^{\circ}$C charge voltage data given by Hunt et al. \cite{Hunt2018}. The value of each parameter discussed above can be found in Table 2 of the supplementary material.

The initial condition is, of course, an idealisation of the expected internal dynamics. Therefore, after the fitting procedure from the first charge and discharge are performed as described in this section, the virtual cell is cycled three times at 0.1C charge and 0.2C discharge before the out-of-sample numerical experiments detailed below were performed. 

The resistance parameters are found by first setting

\begin{equation}
	R = \frac{l}{A\left( \sigma_0 - b\sum_iC_i\right)} = \frac{\alpha}{\beta-\sum_iC_i},
\end{equation}

for

\begin{equation}
	\beta = \gamma \times \max\left(\sum_i C_i \right).
\end{equation}
The parameter $\gamma >1$, ensures the resistance peak is not infinite. The specific value of $\gamma$ controls the sharpness of the resistance curve peak. The parameter $\alpha$ is then obtained by setting 

\begin{equation}
\alpha =R_{max}\times \max\left(\sum_i C_i \right)\times (\gamma-1),
\end{equation}
where $R_{max}$ is the maximum value of the charge resistance from the experimental data. The right hand side of the equation for $\alpha$ is found by matching the peak resistance from the experiment with that of the model. The parameters $\alpha$, $\beta$, and $\gamma$ are re-fit for cells run at different temperatures due to concentration differences. These values may be found in Table 2 in the supplementary material.  

To reflect the conditions of the experiment, the initial condition for the discharge is derived by simulating a full 0.1C charge and using the end-of-charge state. The discharge parameter $k_p^{(T)}$ is the only parameter fit to the discharge voltage data and was found by inspection. The fitting procedure for the 30$^{\circ}$C data is complete once this value is determined. 

To adjust the parameters for the experiments held at different constant temperatures, the value $T$ in the model is altered according to the experiment and the above fitting procedure is performed only for $k_{p/d}^{(T)}$, $k_s^{(T)}$, $S_*^{(T)}$, and $S^{2-}(t=0)$. To ensure the total sulfur mass of all cells was equal, the initial value of precipitated sulfur, $S_p(t=0)$, is set equal to the mass of the sulfur in the 30$^{\circ}$C cell minus the mass of the non-precipitate species. These remaining parameters can be found in Table 3 of the supplementary material.

\section{Model Validation}\label{sec:model_val}

All simulations were performed using the python-based library PyBaMM \cite{Sulzer2020}. Several models from the literature \cite{Marinescu2016, Marinescu2018, Hua2019} have also been implemented along with this paper's model. Therefore, these models are now available to all researchers for immediate use, modification, and direct comparisons \cite{Cornish_LiS-Models_2021}. 

\paragraph{In-Sample Predictions} 

The term $\textit{in-sample}$ refers both to the experimental data set used to fit the model and the set of operational loads used to generate that experimental data. Therefore, model predictions using these operational loads should reproduce the experimental data observed under the same loads. It is possible that no parameter fit can match the model output to the experimental data set with sufficiently low error. Such error indicates that the mathematical form of the model requires modifications, such as additional physics or different formulations of the physics included. 

\begin{figure}[h]
\includegraphics[width=0.5\textwidth]{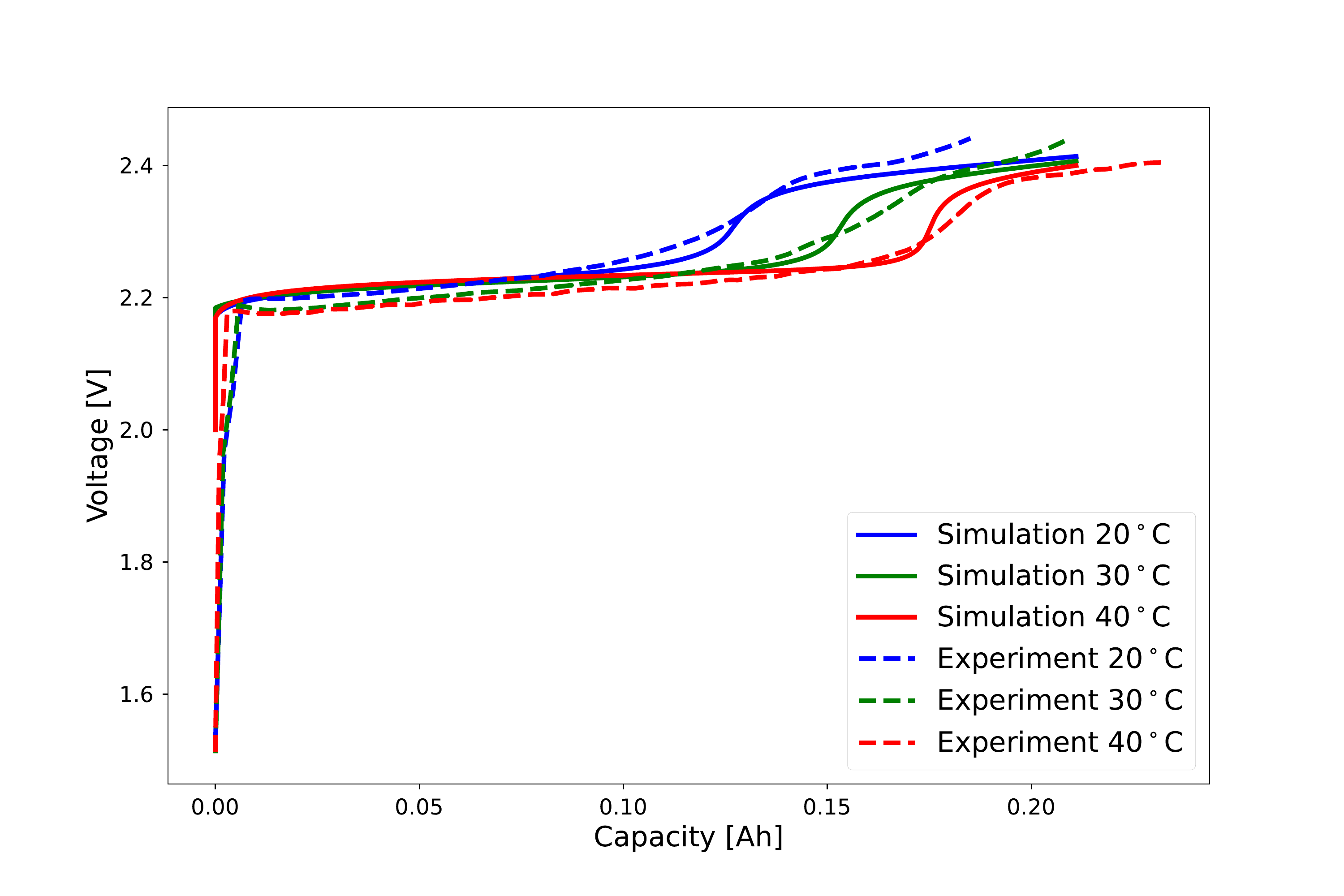}
\caption{In-sample predictions of the 0.1C charge voltage at all three experimental temperatures. Experimental data from Hunt et al. \cite{Hunt2018}. The model is capable of quantitatively capturing this data. }
\label{fig:in_sample_charge_voltage}
\centering
\end{figure}

As shown in Fig.  \ref{fig:in_sample_charge_voltage},  the model achieves numerically accurate values of the charge voltage at all three temperatures. However, the simulations appear to have a sharper transition between the low and high voltage plateaus. It is possible that the sharp transitions between plateaus is due to the formulation of the Butler-Volmer kinetics. Alternative forms of the Butler-Volmer equation include concentration effects which originate from spatial variations of active species near the electrode surface. It is also possible that spatial inhomogeneity across the cathode leads to smoother voltage plateau transitions. The model also deviates slightly from the experimental data at the end of charge. It is unclear from the experimental data whether the end-of-charge voltage spike typically associated with larger currents would be observed at a larger voltage cut-off. This question will be further explored below.  

\begin{figure}[h]
\includegraphics[width=0.5\textwidth]{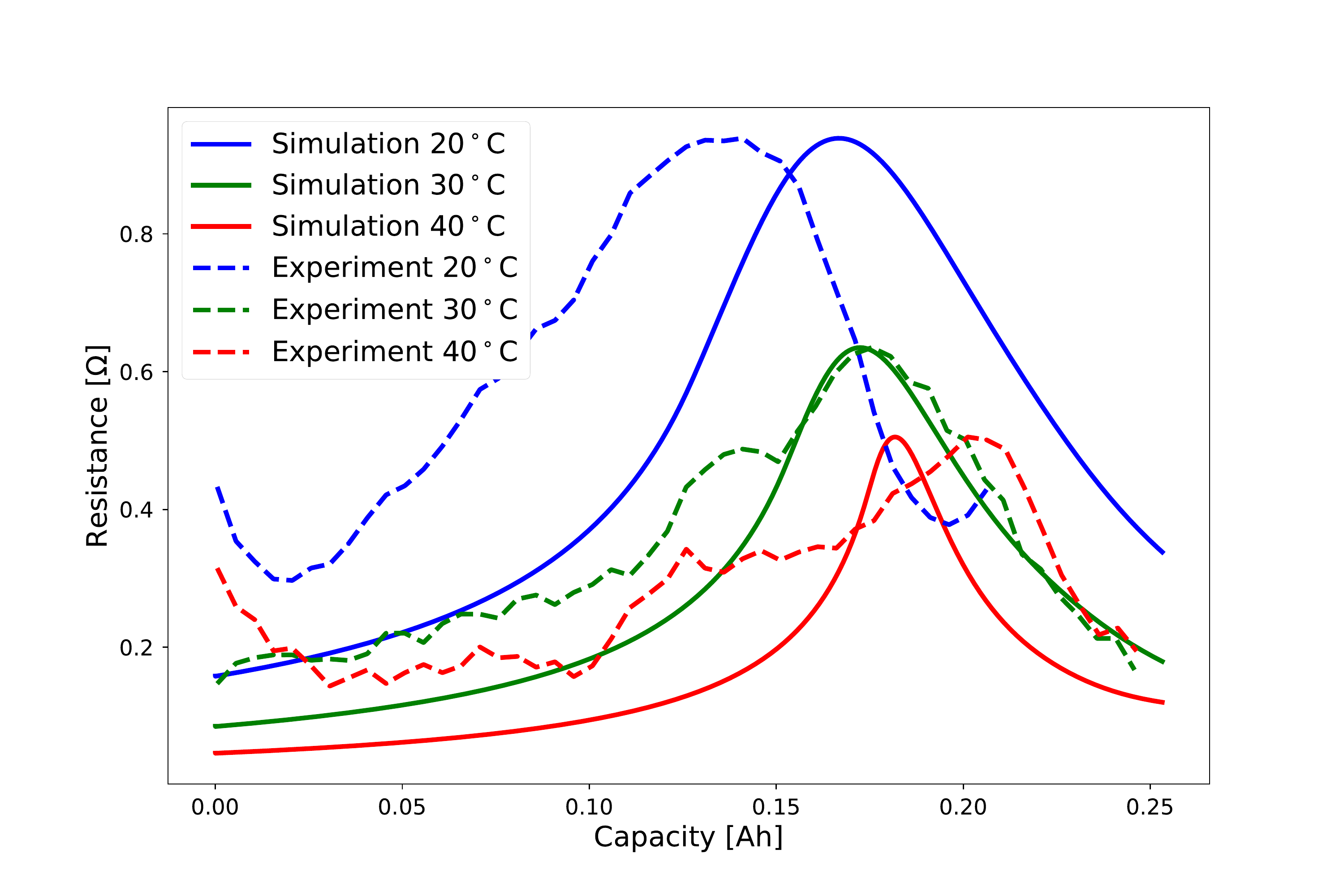}
\caption{In-sample predictions of the 0.1C charge resistance at each experimental temperature. The model is capable of quantitatively retrieving the behaviour from the 30$^{\circ}$C cell while qualitatively retrieving the behaviour from cells held at 20$^{\circ}$C and 40$^{\circ}$C. The discrepancy between model and experimental locations of the resistance peaks may be due to the experimental data gathering method. Experimental data from Hunt et al. \cite{Hunt2018}.}
\label{fig:in_sample_charge_resistance}
\centering
\end{figure}

The charge resistance comparison is shown in Fig. \ref{fig:in_sample_charge_resistance}. The general shape of the simulation appears to match the experimental resistance. However, the location of the simulated resistance peaks does not match that of the experimental resistance peaks for the 20$^{\circ}$C and 40$^{\circ}$C cells. As it is using the hypothesis that resistance is proportional to the concentration of ionic species in the electrolyte, the model predicts that the resistance peak usually occurs in the region of transition between the two plateaus. This misalignment may be due to the ill-defined transition region of the smoothly changing experimental voltage as opposed to the sharp simulation transitions. A notable deviation between the model and the experimental data occurs at the beginning and end of the 20$^{\circ}$C and 40$^{\circ}$C runs. It is unclear why these beginning- and end-of-charge resistance maxima occur in the experimental data, but the resistance model suggests that the electrolyte resistance is not dominant during the initial and final stages of discharge.

\begin{figure}[h]
\includegraphics[width=0.5\textwidth]{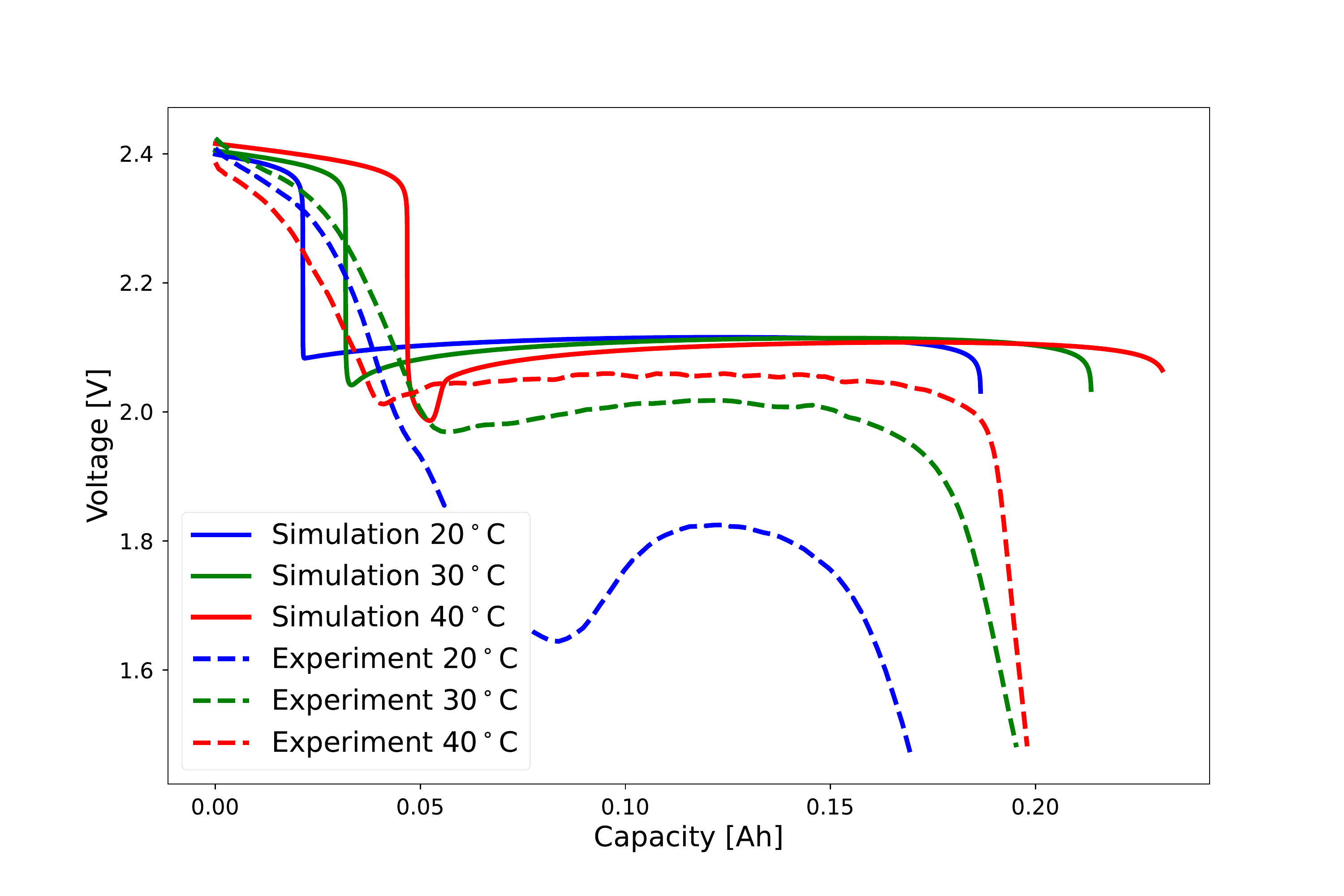}
\caption{In-sample simulations of the 0.2C discharge voltage at each experimental temperature. The model is capable of qualitatively capturing the data from the 30$^{\circ}$C and 40$^{\circ}$C cell but the significant change in behaviour observed for the 20$^{\circ}$C cell is not well captured. Experimental data from Hunt et al. \cite{Hunt2018}.}
\label{fig:in_sample_discharge_voltage}
\centering
\end{figure}

Fig. \ref{fig:in_sample_discharge_voltage} displays the in-sample discharge voltage comparison at various temperatures. Although the simulated discharge voltage curves are qualitatively similar to the experimental data, there are quantitative deviations. The transition regions between high and low voltage plateaus are sharper in the model than in the experiment. It is possible that since the form of the Butler-Volmer equation used for partial currents does not include concentration dependence, the transition between electrochemical processes are sharper than  what is observed experimentally. As in experiments, model predictions show slightly earlier plateau transitions for the higher temperature cell. The relationship between temperature and plateau transition is not currently understood. The quantitative discrepancy between model and simulation may be due to non-electrolyte resistances which our model does not consider. Finally, it appears that there is a significant change in behaviour of the 20$^{\circ}$C cell voltage curve. The voltage plateau transition has a more pronounced dip and the low plateau voltage is not as constant as is the case with higher temperatures. Hunt et al. \cite{Hunt2018} noted that the cause of this low temperature voltage behaviour has not been fully explained. Hunt et al. stated that the behaviour has been associated with cathode preparation techniques which may lead to large transport over-potentials in the cathode. Further, ionic diffusivities and electrolyte conductivity are expected to decrease at lower temperatures. Although some low temperature pouch cell experiments do not yield the same behaviour as shown in Fig. \ref{fig:in_sample_discharge_voltage}, Hunt et al. suggested that multilayer pouch cell experiments may obscure this behaviour due to temperature gradients across the cell layers. Given these explanations, this low temperature voltage behaviour is not captured by the model because the 0D model neither includes transport effects nor the effects of cathode design.  

\paragraph{Out-of-Sample Predictions for Experimental Data}

Out-of-sample data is used to evaluate the accuracy and usefulness of the model. The term out-of-sample is used here to indicate that the experimental data set used to evaluate model accuracy is outside the set of data used to fit the model. Out-of-sample validations ensure that the model can predict cell behaviours outside the set of operations used in the parameterisation data set. Such validation also highlights the boundaries of operational loads for which the model is valid. A model which accurately predicts out-of-sample data can then be used to predict cell behaviour in yet unperformed experiments which are within the operational boundaries. 

\begin{figure}[h]
\includegraphics[width=0.5\textwidth]{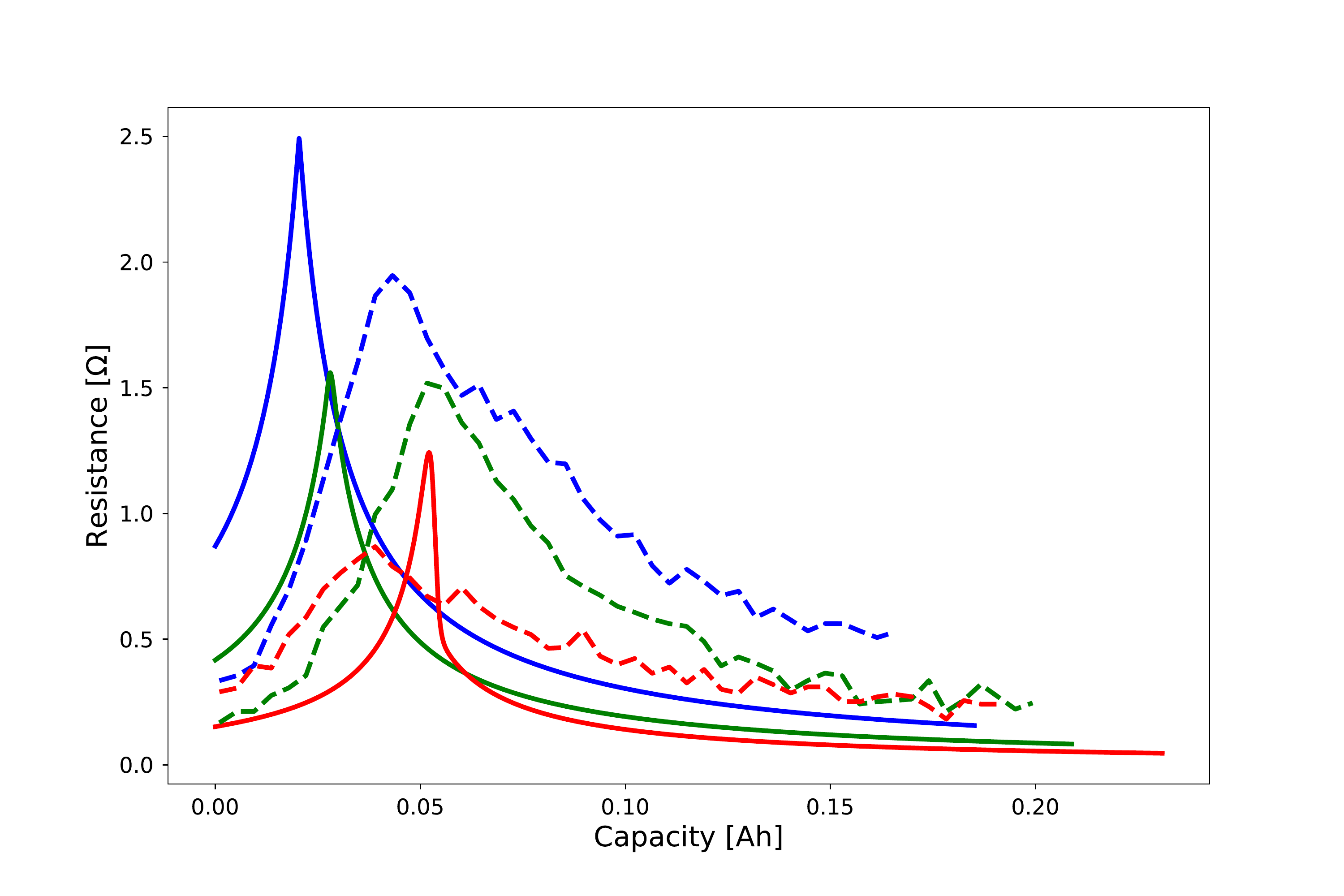}
\caption{Out-of-sample simulations of the 0.2C discharge resistance at each experimental temperature. The model parametrised from charging data is tested against discharge experiments for validation. The model predictions match the experimental data quantitatively for all three cell temperatures. Experimental data from Hunt et al. \cite{Hunt2018}.}
\label{fig:out_of_sample_discharge_resistance}
\centering
\end{figure}

As such, Figure \ref{fig:out_of_sample_discharge_resistance} illustrates the model's capabilities to predict experimental data out-of-sample. Specifically, the experimentally measured Ohmic resistance is compared to the model predictions. The quantitative predictions are not only within the correct order of magnitude but also the model prediction matches the qualitative shape of the resistance curve reasonably well for such a simple model. The model correctly predicts the skewed shape of the resistance profile, where a relatively sharp increase in resistance is followed by a more gradual decrease. Moreover, the peaks in resistance occur at approximately the same location in the experiments. Despite the discrepancy in the accuracy of the discharge voltage, especially for the 20$^{\circ}$C cell, we observe fairly accurate model predictions of resistance values across all temperatures. The model correctly predicts the location of the maximum resistance, as well as the increasing cell resistance with decreasing cell temperature. 

\paragraph{Out-of-Sample Simulations without Experimental Data}

The data reported by Hunt et al. \cite{Hunt2018} does not include all cell behaviour that is relevant for the validation of a LiS model. Therefore, model predictions cannot be quantitatively assessed without implementing the experimental set-up of Hunt et al. with a different set of operational loads and comparing these findings with the out-of-sample model predictions. Alternatively, the model could be refit to data obtained from an experiment with different cells and all operational loads considered in this work. The reporting of model predictions against the widest appropriate set of operating conditions is deemed essential to validating the proposed model. It is also shown that such an analysis is helpful in identifying the most important mechanisms that are missing from this model. In the absence of a complete set of data available from a single cell, a qualitative comparison with the experimentally expected behaviour is presented in this section. 

\begin{figure}[h]
\includegraphics[width=0.5\textwidth]{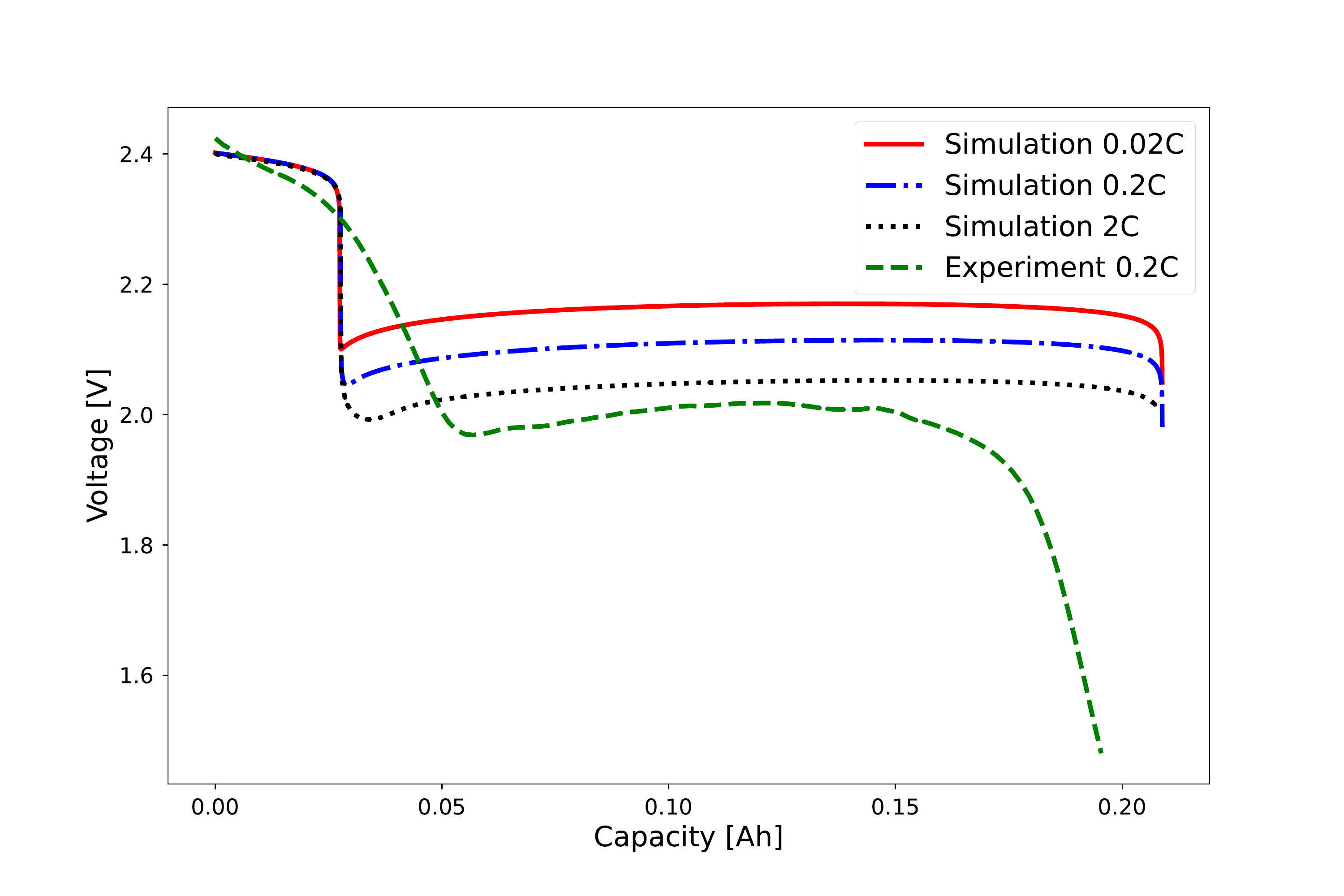}
\caption{Model predictions of the effect of discharge rate voltage for a 30$^{\circ}$C cell. Li-S cells typically increase lower plateau over-potential with decreasing current, as is observed the model predictions. Experimental data from Hunt et al. \cite{Hunt2018}.}
\label{fig:discharge_current_effects}
\centering
\end{figure}

As summarised in Table \ref{tab:tests}, it is common to observe that decreasing the discharge current leads to an increase in over-potential and an increase in capacity primarily in the low plateau. Model predictions of the cell voltage during constant current discharge at three current magnitudes displayed in Fig. \ref{fig:discharge_current_effects} show that the model predicts higher over-potential in the lower plateau with a decrease in discharge current. The upper plateau voltage and capacity is mostly unaffected. However, the expected low plateau capacity change is not observed. The dependence of the discharge capacity on current is possibly a spatial effect \cite{Zhang2016, Zhang2018} and therefore a zero-dimensional is not expected to produce such behaviour. Alternatively, it has been hypothesised that the capacity-rate effect is due to active surface area passivation through rate dependent precipitation dynamics \cite{Ren2016, Andrei2018, Xiong2019, Danner2019}. Precipitation in our model does not effect active surface area and is not rate-dependent. The absence of such dynamics could result in the discrepancy between model predictions and expected cell behaviour. Finally, it should also be noted that the initial value of the discharge voltage of approximately 2.41 V remains roughly the same across to different discharge currents. It is common to observe similar initial voltage values across C-rates, but is not true for all models \cite{Fronczek2013}.

\begin{figure}[h]
\includegraphics[width=0.5\textwidth]{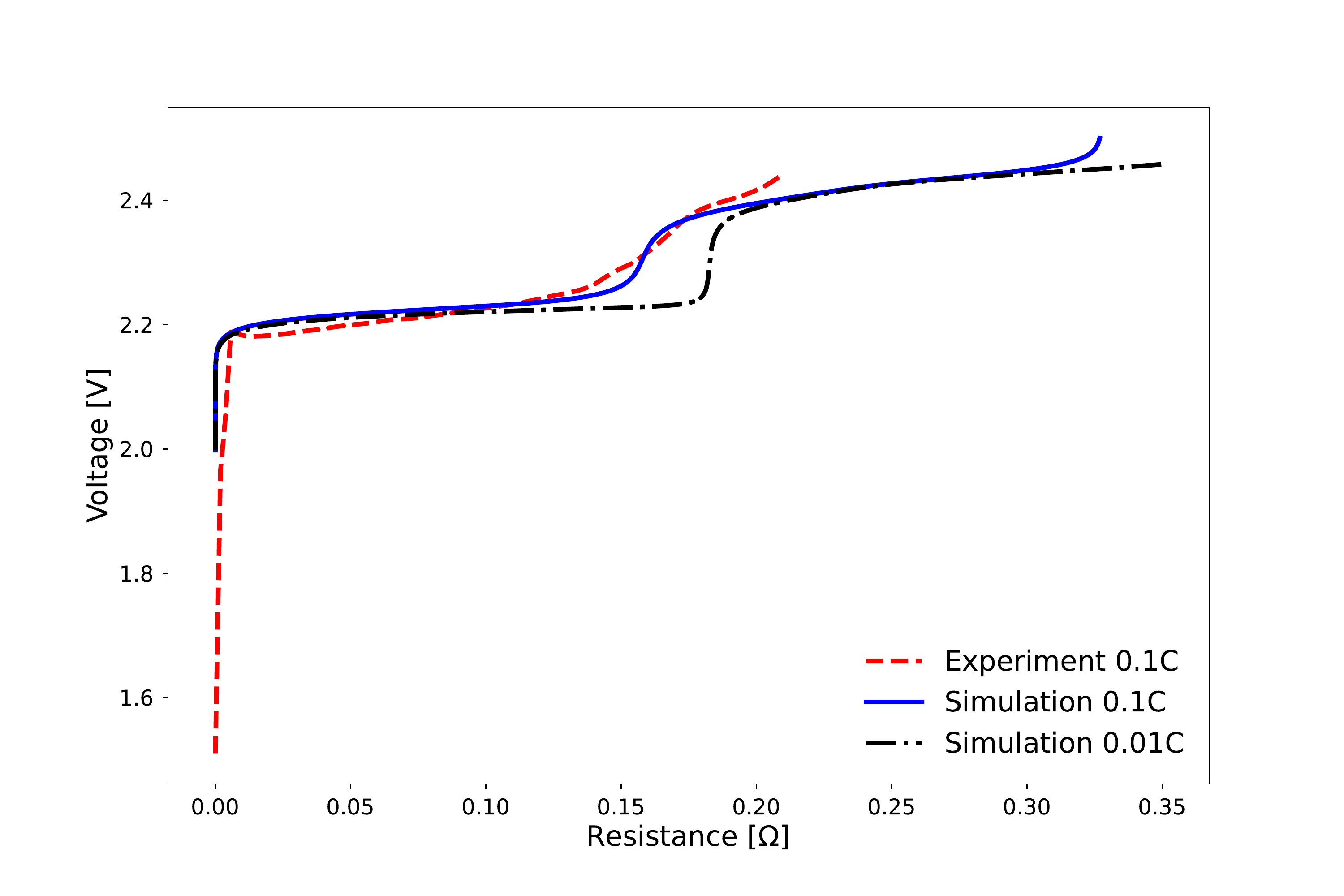}
\caption{Simulations of the effect of charge current on voltage for a 30$^{\circ}$C cell. At low currents, Li-S cells are expected to have a seemingly infinite charge-capacity in the upper plateau. Larger currents lead to an upper-plateau spike in voltage above operating bounds. The model captures this relationship between current and end-of-charge voltage dynamics. Experimental data from Hunt et al. \cite{Hunt2018}.}
\label{fig:charge_current_effects}
\centering
\end{figure}

Decreasing the charge current typically leads to the infinite-capacity phenomenon associated with the polysulfide shuttle. There appears to be a critical current above which the shuttle effect is not dominant, leading to an end-of-charge voltage spike. The polysulfide shuttle is a spatial phenomenon and therefore a zero-dimensional model is not expected to capture this. Nonetheless, our model utilises a pseudo-spatial mechanism to capture the shuttle effect. Fig. \ref{fig:charge_current_effects} demonstrates the end-of-charge voltage bifurcation. It is unclear if the experimental voltage data would increase or level-off if charged longer. In any case, the model indicates the voltages separate at a larger charge capacity than the experiment considered. Finally, as in the discharge-current test, we note initial voltage stability with respect to charge currents. 

\begin{figure}[h]
\includegraphics[width=0.5\textwidth]{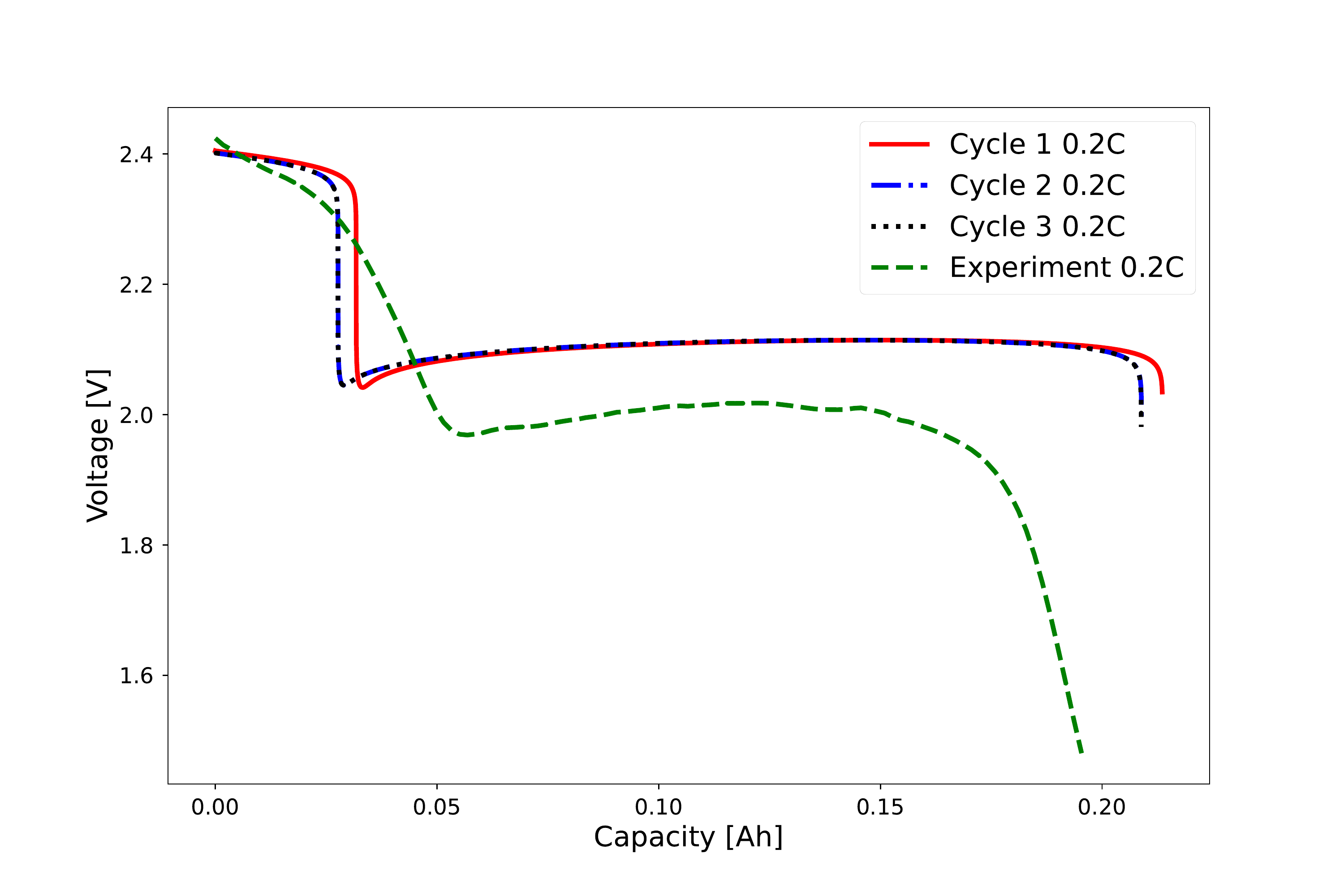}
\caption{Model predictions of the effect of cycling with 0.2C discharge and 0.1C at 30$^{\circ}$C. Li-S cells typically reduce capacity with increased cycling. The model captures this capacity fade after the first cycle but then settles on a fixed discharge-charge path. Experimental data from Hunt et al. \cite{Hunt2018}.}
\label{fig:discharge_cycle_effects}
\centering
\end{figure}

Despite charge and discharge behaviour, both in terms of cell voltage and cell resistance, another common feature of Li-S cells is their capacity fade with cycling. While the model proposed does not include degradation, it can retrieve memory effects due to precipitation/dissolution dynamics. Model predictions for multiple cycles are reported in Fig. \ref{fig:discharge_cycle_effects}. To coincide with the experimental data, the model was cycled with 0.1C charge and 0.2C charge and were time-limited. The cell voltage curves evolve only within the first two cycles. 

Although this is an idealistic virtual cell, there are several features of this test which are instructive. Experimental cells should not be expected to be identical in terms of micro-states, such as mass distribution of species. For commercial cells, such as those used in Hunt et al. \cite{Hunt2018}, it is not common to observe significantly different macroscopic measurements across cells. Therefore, these (possibly small) differences between cell micro-states only lead to small differences in macroscopic measurements. Voltage curves across different cells are similar and changing the discharge current does not significantly change the initial voltage value. Therefore, macroscopic model predictions should not be highly sensitive to micro-states. After the precipitation/dissolution memory effects alter the first discharge voltage curve, variations in the initial mass distribution of the model does not lead to significant changes in voltage behaviours, as expected for experimental cells. 

The model's internal state at the start of discharge between the first and second cycle changes, while the difference between the second and third cycle is insignificant. This suggests that without a degradation mechanism the model has reached a steady state with respect to the cycling procedure. Although charge and discharge are asymmetric processes, the path through state space appears to be a closed loop. We also observe that the limited cycling which we have performed reduces the capacity and voltage. This is consistent with experimental data. Judging from the published literature, the community has not yet agreed upon the specific mechanism which leads to capacity fade across charge/discharge cycles. Consequently, there is not an agreed upon model for capacity fade. In any case, the proposed model does not appear to capture this behaviour for the operational conditions trialled.

\begin{figure}[h]
\includegraphics[width=0.5\textwidth]{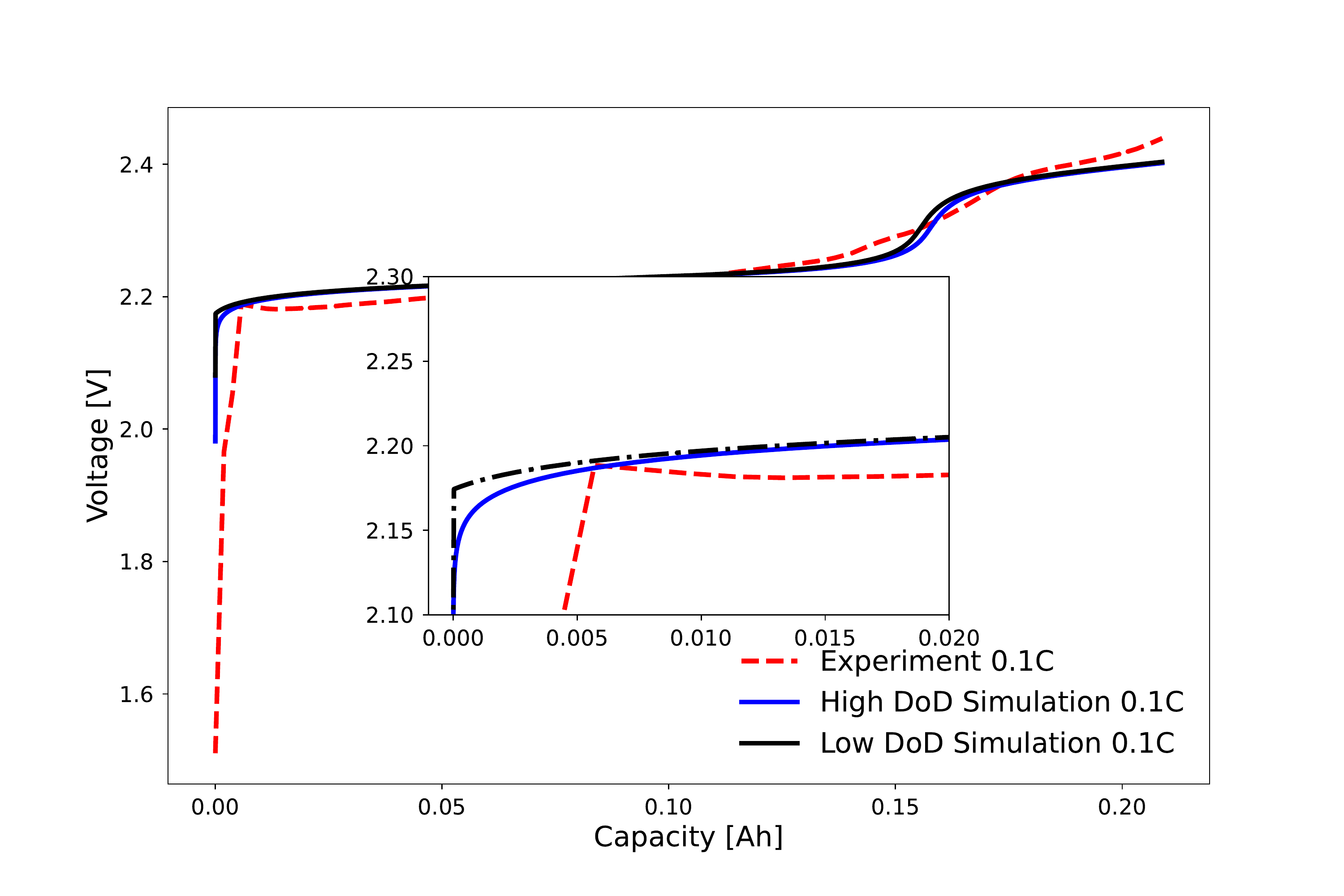}
\caption{Model predictions of the effect of depth-of-discharge on the initial stages of charge voltage for a 30$^{\circ}$C cell. It is common to observe a small kink in the charge voltage curve prior to the start of the lower plateau depending on the depth-of-discharge of the previous discharge. The proposed model does not produce such a relationship. Experimental data from Hunt et al. \cite{Hunt2018}.}
\label{fig:DoD_effects}
\centering
\end{figure}

The final out-of-sample prediction compared against typical Li-S behaviours is the relationship between the depth-of-discharge and the voltage kink observed during the initial stages of charge. The depth-of-discharge (DoD) was determined by the end-of-discharge voltage. A high DoD corresponds to $V = 2.08$ while a low DoD corresponds to $V=1.98$. The charge voltage simulations displayed in Fig. \ref{fig:DoD_effects} are nearly identical whether starting from a relatively high or low depth-of-discharge. The simulated voltage curve changes more dramatically when the previous discharge stops during the low plateau rather than when the end of discharge is implemented during the voltage drop-off period afterwards (not shown). However, the sharp drop in voltage at the end of discharge is primarily due to the high sensitivity of the discharge voltage to the end-of-discharge internal state. Therefore, during the end-of-discharge voltage drop, a larger DoD as measured by voltage equates to only a slight variation in the internal state. Since the charge voltage curve is stable with respect to these small changes in the internal state, the model does not predict noticeable differences in the charge voltage curve. If the end-of-charge voltage drop corresponded to more significant changes in the internal state then the charge voltage would be more strongly dependent upon DoD. Such a dependence may be incorporated through porosity effects, a concentration dependent Butler-Volmer equation, or a spatially resolved model. 

\section{Conclusion}\label{sec:conclusion}

A single Li-S cell held at a constant temperature was modelled with a zero-dimensional model. The model utilises a three-stage electrochemical reaction as well as a shuttle effect. Resistance is modelled by using the concentration dependent function developed by Zhang et al. \cite{Zhang2016}. As Zhang et al. observed, even spatially resolved models do not correctly predict experimental resistance data even qualitatively. Zhang et al. showed that a zero-dimensional model can achieve the correct form of discharge resistance data by using the concentration dependent resistance function. However, their model was not able to correctly predict charge dynamics. On the other hand, the simpler model of Marinescu et al. \cite{Marinescu2016} allowed charge and discharge dynamics to be qualitatively modelled, but could not predict qualitatively correct resistance curves due to the electrochemical pathway. The new model proposed combines the functionality of all previous zero-dimensional models, while keeping mathematical complexity to a minimum.

First, the model quantitatively fits the charge voltage curve and charge resistance curve. The discharge voltage predictions are less accurate quantitatively, but still yield a good fit. Second, the model accurately predicts discharge resistance data. The data used to fit the model does not include the discharge resistance data, therefore this test represents a powerful validation of the model. Third, the typical Li-S cell behaviour observed in the literature and summarised in Table \ref{tab:tests} were considered qualitatively and out-of-sample. The relationship between discharge voltage and current was predicted with reasonable accuracy considering the model simplicity. The relationship between charge voltage and current was captured, albeit the bifurcation due to shuttle is observed in the simulation at a larger capacity than in the experimental cell. Model errors are possibly due to mass transport limitations and remain to be tested in one dimensional models. Capacity fade with cycling is also tested. Model predictions settle onto a consistent charge-discharge cycle. Finally, the relationship between the depth-of-discharge and initial charge voltage kink was explored but without observing the phenomenon during simulation. The proposed model accurately predicts many Li-S cell behaviours both quantitatively and qualitatively but will require additional physics to reduce model error.  

The proposed model, along with those in \cite{Marinescu2016, Marinescu2018, Hua2019}, were implemented using the open-source library PyBaMM \cite{Sulzer2020} and are available to everyone in an effort towards transparency, rigour, and open-access science \cite{Cornish_LiS-Models_2021}, (section 14 of \cite{Robinson2021}).

Along with a model, this work proposed a minimum test set for cell-level physics-based models of Li-S battery technology. The test set included operational loads which yield typical behaviour for macroscopic variables such as voltage and resistance. Model testing using this test set will allow the assessment of model performance to be made more easily and rigorously. Each model represents a competing hypothesis for a mechanistic explanation of Li-S behaviour. Each hypothesis requires both evidence that model predictions match available data as well as evidence that alternative hypotheses do not. The standard test set ensures both strands of evidence will be available. Ideally, competing models should be validated with the same data set. In the absence of an open data set, we encourage all researchers to simulate this set of common behaviours for Li-S cells to allow the research community to make more informed model selections and modifications. Rigorous validation and selection will produce models which both accurately predict cell behaviour under unseen operational loads and provide mechanistic explanations for such behaviour. The former is essential for battery management systems while the latter will help optimise the development cycle of Li-S technology.

 \section*{Acknowledgments}
 The authors would like to acknowledge financial support from the EPSRC Faraday Institution Lithium Sulfur Batteries Project (EP/T012404/1, Grant Number FIRG014)

\section*{References}

\bibliography{mybibfile}

\end{document}